# RADIATIVE n$^{11}$B CAPTURE ACCOUNTING 21 AND 430 keV RESONANCES

S. B. Dubovichenko[1]

[1] *Fessenkov Astrophysical Institute "NCSRT" NSA RK, 050020, Almaty, Kazakhstan*

N.A. Burkova [2, *, †]

[2] *Al-Farabi Kazakh National University 050000, Almaty, Kazakhstan*
*natali.burkova@gmail.com

In the framework of the modified potential cluster model the possibility of describing the available experimental data for the total cross sections for $n^{11}$B radiative capture at thermal and astrophysical energies were considered with taking into account the 21 and 430 keV resonances.

*Keywords*: radiative capture; potential cluster model; resonances.

**PACS:** Number(s): 21.60.Gx, 25.20.-x, 25.40.Lx, 26.35.+c, 26.

## 1. Introduction

During the last near 50 years successful development of the theoretical nuclear physics was provided to a large extent by some microscopic models known as Resonating Group Methods (RGM), [1, 2] Generating Coordinate Method (GCM), [3] and algebraic version of RGM [4]. However, the possibilities of the simple potential cluster models (PCM) [5] basing on the two-body potentials with the Pauli forbidden states (FS) are not exhausted yet. When in PCM the resonating behavior of elastic scattering phase shifts at low energies [6-10] is included, then we arrive to the modified potential cluster model (MPCM).

Not always rather complicated calculations in RGM are the only ones for the explanation of the experimental data on the radiative capture reactions at low and ultra-low energies. As we showed earlier [6, 10] MPCM including the orbital states classification according to the Young diagrams, make it possible to reproduce the data on the total cross sections of the thermonuclear radiative capture processes on light nuclei and even provides some predictive proposals.

Therefore, continuing the study of the reactions of neutrons radiative processes by light nuclei within the MPCM [6, 10] we are reporting on the $n^{11}$B → $γ^{12}$B reaction at thermal and astrophysical energies. As there are no data on the elastic scattering phase shifts in initial channel, the structure of the resonance states of final nucleus $^{12}$B formed the basis for the construction of the two-body interaction potentials for continuous scattering states.

Final bound or ground states (BS or GS) of $^{12}$B generated in the treating reaction have the same cluster configuration as the initial ones. So, the two-body potentials have constructed on the basis of the description of these particles binding energy in the final nuclei and some basic characteristics of these states, for example, the binding energy and its asymptotic constants (AC) [10].

The treating process $n^{11}$B → $γ^{12}$B may play a definite role in some models of Big Bang [11-15]. In such models the primary nucleosynthesis is assumed to develop according the following chain of nuclear reactions

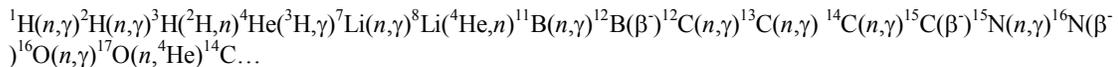

$^1$H($n,γ$)$^2$H($n,γ$)$^3$H($^2$H,$n$)$^4$He($^3$H,$γ$)$^7$Li($n,γ$)$^8$Li($^4$He,$n$)$^{11}$B($n,γ$)$^{12}$B($β^-$)$^{12}$C($n,γ$)$^{13}$C($n,γ$) $^{14}$C($n,γ$)$^{15}$C($β^-$)$^{15}$N($n,γ$)$^{16}$N($β^-$)$^{16}$O($n,γ$)$^{17}$O($n,^4$He)$^{14}$C…

Moreover, it is not except that such results may depend on the dark energy state [16], on the baryonic matter perturbations' growing [17] or on the rate of early Universe rotation [18]. In doing this the plasma's perturbations may as stimulate the nucleosynthesis process, [19] as far as depress it due to non-baryonic matter perturbations' growth [20] or to the cosmic strings oscillations [21].

## 2. Structure of states in *n*$^{11}$B system

In view of absence of the complete tables for the product s of Young schemes for the systems with a number of nucleons more than eight, [22] the presented below results might be consider as the qualitative estimation of the possible orbital symmetries in the ground state of $^{12}$B in the single taken $n^{11}$B cluster channel.

At the same time, just basing on such classification we succeeded to reproduce, and, what is more important, to explain available experimental data on the radiative nucleon capture reactions in channels $p^{13}$C, [23] $n^{14}$C, and $n^{14}$N, [24] as well as a greater number of neutron radiative capture processes[10, 25]. So, it is reasonable to apply our approach basing on the classification of cluster states by orbital symmetry what leads to appearing of a number of FS and allowed states (AS) in partial two-body potentials, and as a consequence relative motion wave functions (in present case those refer to the neutron and $^{11}$B system) have a set of nodes.

Let us assume the {443} orbital Young scheme for $^{11}$B, then within the 1p-shell treating of $n^{11}$B system one has {1} × {443} → {543} + {444} + {4431} [22, 26]. The first scheme in this product is compatible with the orbital momenta $L$ = 1, 2, 3, 4, it is forbidden, as no more than four nucleons may be on s-shell. Second scheme is allowed and compatible with the orbital momenta $L$ = 0, 2, 4, as for the third one also allowed the corresponding momenta are $L$ = 1, 2, 3 [26].

It should be noted that even such qualitative analysis of orbital symmetries allow to define that there are forbidden states in $P$- and $D$ – waves, and no ones in $^3S_1$ – state (here the notation $^{(2S+1)}L_J$ is used). Actually, such a structure of FS and AS allows to construct two-body interaction potentials required for the calculations of total cross sections for the treating reaction.

Thereby, restricting the problem by the lower partial waves with orbital momenta $L$ = 0, 1 and 2 can be said that for $n^{11}$B system ($^{11}$B in GS has quantum numbers $J^\pi$, $T$ = 3/2$^-$, 1/2 [27]) interactive potential for $^3S_1$ – wave should include only AS, as far as potentials for $^3P$- waves they should have both AS and FS. Allowed state, namely $^3P_1$, corresponds to the GS of $^{12}$B with quantum numbers $J^\pi$, $T$ = 1$^+$, 1 with binding energy –3.370 MeV in $n^{11}$B channel [27]. Generally speaking, some scattering states in $n^{11}$B channel, as well as BS, may be mixed by channel spin $S$ = 1 or 2, but here we are assuming the spin states as pure triplet.

Let us list now excited but bound states in $^{12}$B nuclei in $n^{11}$B- channel [27]. Notice, after the level energy relatively the GS, the energy relatively the threshold of $n^{11}$B channel is pointed in brackets.

1. First excited state with $J^\pi$ = 2$^+$; 0.95 MeV (-2.4169 MeV). It can be associated with triplet $^3P_2$ – wave with FS.

2. Second excited state with $J^\pi$ = 2$^-$; 1.67 MeV (-1.6964 MeV). It can be associated with triplet $^3D_2$ – wave with FS.

3. Third excited state with $J^\pi$ = 1$^-$; 2.62 MeV (-0.7492 MeV). It can be associated with triplet $^3S_1$ – wave without FS.

4. Fourth excited state with $J^\pi$ = 0$^+$; 2.72 MeV (-0.647 MeV). It can be associated with triplet $^3P_0$ – wave with FS.

Notice, if the bound state has a binding energy less than 1 MeV, then one may neglect the most strong $E1$ transition, cause of its minor input to the total cross section. Consider now spectrum of resonance states (RS) in $n^{11}$B system, [27] i.e. states with positive energies.

1. First RS at 20.8(5) keV with $J^\pi$ = 3$^-$ and less than 1.4 keV width. It can be associated with $^3D_3$ scattering wave with bound FS.

2. Second RS at 430(10) keV with $J^\pi$ = 2$^+$ and 37(5) keV width. It can be associated with $^3P_2$ scattering wave with bound FS.

3. Third RS at 1027(11) keV with $J^\pi$ = 1$^-$ and 9(4) keV width. It can be associated with $^3S_1$ scattering wave without bound FS.

The third RS and following after resonances lay at higher than 1 MeV energies, so the not be considered. Below 1 MeV here are no resonance states which may be associated with $^3S_1$ scattering wave [27]. So, its phase shifts are close or equal zero, and as there is no FS in this wave, the interaction potential may be regarded zero [10].

Thus, we will consider the following electromagnetic transitions to the ground state and four excited states (ES):

(a) $^3S_1 \xrightarrow{E1} {}^3P_1$  (g) $^3S_1 \xrightarrow{E1} {}^3P_0$

(b) $^3D_3 \xrightarrow{E1} {}^3P_2$  (h) $^3P_2 \xrightarrow{E1} {}^3D_2$

(c1) $^3P_2 \xrightarrow{E1} {}^3S_1$  (i) $^3P_2 \xrightarrow{M1} {}^3P_1$

(c2) $^3P_1 \xrightarrow{E1} {}^3S_1$  (j) $^3P_2 \xrightarrow{M1} {}^3P_2$

(c3) $^3P_0 \xrightarrow{E1} {}^3S_1$  (k) $^3S_1 \xrightarrow{M1} {}^3S_1$

(f) $^3S_1 \xrightarrow{E1} {}^3P_2$  (l) $^3D_3 \xrightarrow{M1} {}^3D_2$

Detailed discussion on each transition will be given in Section 5.

## 3. Calculating methods

Expression for calculating of the radiative capture process total cross-sections σ($NJ$, $J_f$) for $EJ$ and $MJ$ transitions in the cluster model has the following form [6, 28-30]

$$\sigma_c(NJ, J_f) = \frac{8\pi Ke^2}{\hbar^2 q^3} \frac{\mu}{(2S_1+1)(2S_2+1)} \frac{J+1}{J[(2J+1)!!]^2} \times A_J^2(NJ, K) \sum_{L_i, J_i} P_J^2(NJ, J_f, J_i) I_J^2(J_f, J_i)$$

where $\sigma$ is radiative capture total cross section; $\mu$, $q$ are the reduced mass and the wave number of particles in the initial channel; $K$, $J$ are the wave number and the moment of $\gamma$ in the final channel; $N$ is $E$ or $M$ transitions of multipolarity J from the $J_i$ initial to $J_f$ final state of a nucleus.

For the electric orbital $EJ(L)$ transitions $P_J$, $A_J$, and $I_J$ are of the following form [6, 28]

$$P_J^2(EJ, J_f, J_i) = \delta_{S_i S_f} \left[ (2J+1)(2L_i+1)(2J_i+1)(2J_f+1) \right] (L_i 0 J 0 | L_f 0)^2 \begin{Bmatrix} L_i & S & J_i \\ J_f & J & L_f \end{Bmatrix}^2,$$

$$A_J(EJ, K) = K^J \mu^J \left( \frac{Z_1}{m_1^J} + (-1)^J \frac{Z_2}{m_2^J} \right), \quad I_J(J_f, J_i) = \langle \chi_f | R^J | \chi_i \rangle.$$

Here $S_i$, $S_f$, $L_f$, $L_i$, $J_f$, $J_i$ are the spins, orbital and total momenta of particles in the initial channel (*i*) and the nucleus in the final channel (*f*) correspondingly; $m_1$, $m_2$, $Z_1$, $Z_2$ are the masses and charges of particles in the initial channel; $I_J$ is the integral over the initial $\chi_i$ and final $\chi_f$ states wave functions, i.e. over the clusters relative motion functions with the intercluster distance $R$,

For the treating of dipole magnetic $M1(S)$-transition of spin type [31] the following expressions have been used [6, 10]

$$P_1^2(M1, J_f, J_i) = \delta_{S_i S_f} \delta_{L_i L_f} \left[ S(S+1)(2S+1)(2J_i+1)(2J_f+1) \right] \begin{Bmatrix} S & L & J_i \\ J_f & 1 & S \end{Bmatrix}^2,$$

$$A_1(M1, K) = i \frac{e\hbar}{m_0 c} K\sqrt{3} \left[ \mu_1 \frac{m_2}{m} - \mu_2 \frac{m_1}{m} \right], \quad I_J(J_f, J_i) = \langle \chi_f | R^{J-1} | \chi_i \rangle, \quad J = 1,$$

where $m$ is nuclear mass; $\mu_1$ and $\mu_2$ magnetic momenta of clusters [32, 33].

For the methods developed for the two-body cluster potentials for the partial waves $L$ see Refs. 10 and 34. In presented calculations the following values for particle masses have been used $m_n$ = 1.00866491597 a.m.u. [32] and $m(^{11}B)$ = 11.0093052 a.m.u. [35], and the value of $\hbar^2/m_0$ constant was taken equal to 41.4686 MeV·Fm$^2$.

## 4. Interaction potentials

Partial $n^{11}B$ interaction potential for each orbital momentum $L$, total momentum $J$, and parity $\pi$ as usual is presented in the Gaussian form

$$V(J^\pi, L, r) = -V_0(^{2S+1}L_J) \exp[-\alpha(^{2S+1}L_J) r^2], \quad (1)$$

For the potential of resonating $^3P_2$ – wave with one BS, which is forbidden, the following parameters have been obtained basing on the $^{12}B$ spectra as well as data on the elastic scattering in n + $^{11}B$ channel

$$V_0(^3P_2) = 11806.017 \text{ MeV}, \quad \alpha(^3P_2) = 15.0 \text{ Fm}^{-2}. \quad (2)$$

Within this potential the resonance energy level $E$ = 430 keV and its width 37 keV according Ref. 29 data have been reproduced exactly: at level energy phase shift turned to be 90.0°(1). To calculate the level width basing on the phase shift $\delta$ we used expression $\Gamma = 2(d\delta/dE)^{-1}$. Energy dependence of $^3P_2$ phase shift is given on Fig. 1.

Let us note, the analysis of resonance scattering bellow 1 MeV when the resonance width is about 10 – 50 keV, and number of BS is given, the interaction potential may be defined completely unambiguously. At given number of BS the potential depth is fixed by the resonance energy of the level, and its width is defined by the resonance width.

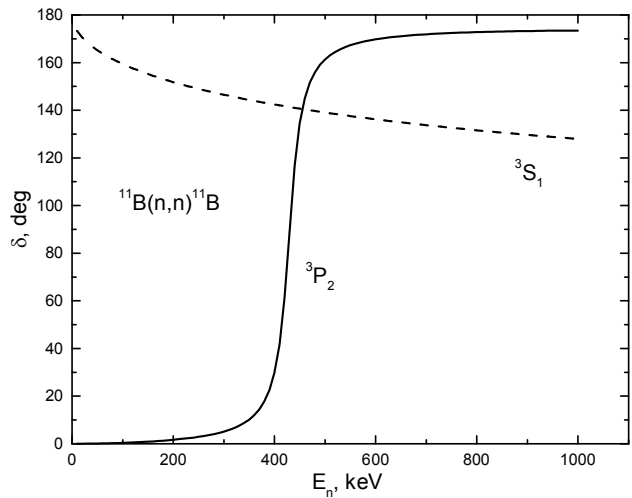

Fig 1. Elastic scattering $^3P_2$ phase shift with resonance at 430 keV (solid curve); $^3S_1$ phase shift (dash).

The parameters error usually not more than accuracy of the definition of the corresponding level and it is about 3 ÷ 5%. Same remark refers to the reconstruction procedure of the partial potentials using scattering phase shifts and information of the resonances in spectrum of final nucleus [10, 34]. However some ambiguity in such potential may exist.

Basing on the given above classification of states in treating system by Young schemes the number of AS and FS may be defined only, but definite conclusion either AS is bound or not cannot be done. In particular, AS in $^3P_2$ scattering wave might not be necessarily bound. So, we consider all FS in scattering waves as bound, what allows exclude the pointed ambiguity.

For the none-resonating $^3P_0$ и $^3P_1$ scattering waves with bound FS we propose a following set of parameters

$$V_0(^3P_0, ^3P_1) = 11000.0 \text{ MeV}, \quad \alpha = 15.0 \text{ Fm}^{-2}. \tag{3}$$

This potential gives scattering phase shifts less than 0.5° in the energy region bellow 1.0 MeV.

For the resonating $^3D_3$ wave with FS the following set of potential parameters was found

$$V_0(^3D_3) = 129.305 \text{ MeV}, \quad \alpha = 0.1 \text{ Fm}^{-2}. \tag{4}$$

Fig. 2 shows corresponding phase shifts with resonance at 20.9(1) keV and width less than 1 keV, what is in excellent coincidence with known data [27].

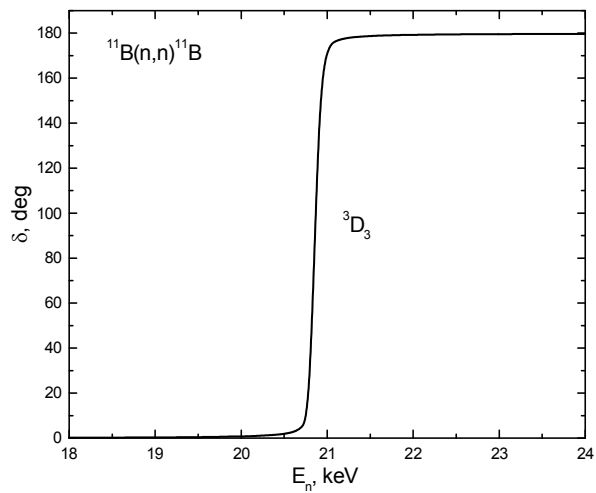

Fig 2. Elastic scattering $^3D_3$ phase shift with resonance at 20.9(1) keV.

One should keep in mind, that if potential contains $N + M$ forbidden and allowed states, then it obeys to the generic Levinson theorem, and the corresponding phase shifts at zero energy should start from $\pi \cdot (N + M)$ [5]. However, on Figs. 1 and 2 for more conventional representation, the results for P and D phase shifts where bound FS appear are given from zero, but not from 180°. As far as S phase shift which has bound AS is given from 180° according Levinson theorem.

The parameters set for the none-resonating $^3D_2$ and $^3D_1$ waves with FS was used

$$V_0(^3D_2, {}^3D_1) = 78.0 \text{ MeV}, \ \alpha = 0.1 \text{ Fm}^{-2}. \tag{5}$$

It led to the values of phase shifts less than 0.1° in the energy region bellow 1.0 MeV.

Now, let us turn to the description of the constructed for BS potentials, both for the GS, and all excited but bound states in $n^{11}$B channel. For the ground $^3P_1$ state of $^{12}$B in $n^{11}$B channel with FS we found

$$V_{BS}(^3P_1) = 3183.6365 \text{ MeV}, \ \alpha = 4.0 \text{ Fm}^{-2}. \tag{6}$$

This potential allows to obtain the mass radius $R_m$ = 2.36 Fm, charge radius $R_{ch}$ = 2.41 Fm, calculated binding energy -3.3700 MeV [36] coincides with experimental value -3.370 MeV [27]. For the asymptotic constant (AC) written in the dimensionless form (see Ref. 37)

$$\chi_L(r) = \sqrt{2k_0} C_w W_{-\eta L+1/2}(2k_0 r)$$

value 0.43(1) was obtained over the range 2 – 14 Fm. Error of the calculated constant is defined by its averaging over the mentioned above distance range. For the mass and charge radii of $^{11}$B value 2.406(29) Fm was used [38]. It seems $^{12}$B radius should not differ greatly from those of $^{11}$B and $^{12}$C, for the latter it is also known and equals 2.4702(22) Fm [38]; neutron mass radius was taken same as the proton one 0.8775(51) Fm [32]. Note, in scattering channel the corresponding phase shift calculated with potential (6) smoothly slows down to 178° at 1.0 MeV.

For the AC of $^{12}$B in GS of $n^{11}$B cluster channel numerical value 0.245 Fm$^{-1}$ (or 0.495 Fm$^{-1/2}$) [39] was obtained with accounting of identity of the nucleons (see Eq. 83b in Ref. 40). In Ref. 39 another definition of the AC was used

$$\chi_L(r) = C \cdot W_{-\eta L+1/2}(2k_0 r),$$

what differs from the used here by a factor $\sqrt{2k_0}$ equals 0.88 Fm$^{-1/2}$ for the GS, what gives $C_w$ = 0.56.

We suggest one more set of potential parameters for the GS which reproduces given above AC

$$V_{BS}(^3P_1) = 1606.331 \text{ MeV}, \ \alpha = 2.0 \text{ Fm}^{-2}. \tag{7}$$

Within this potential we obtained AC = 0.55(1) over the range 2 – 16 Fm, charge and mass radii 2.37 Fm and 2.41 Fm at the binding energy -3.3700 MeV obtained with 10$^{-4}$ MeV accuracy [36]. Scattering phase shift shows same behavior as in case of potential (6).

For the parameters of $^3P_2$ potential with FS corresponding to the first ES of $^{12}$B nucleus ($J^\pi = 2^+$) in $n^{11}$B channel lying at 0.95 MeV the following values have been found

$$V_{BS}(^3P_2) = 3174.75797 \text{ MeV and } \alpha = 4.0 \text{ Fm}^{-2}. \tag{8}$$

This potential gives binding energy -2.4169 MeV with accuracy $\varepsilon$ = 10$^{-4}$ quite well coinciding with experimental value -2.41686 MeV, [27] charge radius 2.41 Fm and AC = 0.38(1) over the range 2 – 14 Fm. Ref. 39 reported AC values 0.098 Fm$^{-1}$ or 0.313 Fm$^{-1/2}$, and recalculation to the dimensionless AC with $\sqrt{2k_0}$ = 0.81 Fm$^{-1/2}$ it turned to be equal 0.386 what is in agreement with AC obtained with potential (8). Corresponding phase shift smoothly decreases up to 177° at 1.0 MeV.

For the parameters of $^3D_2$ potential with FS corresponding to the second ES of $^{12}$B nucleus ($J^\pi = 2^-$) in $n^{11}$B channel lying at 1.67 MeV the following values have been found

$$V_{BS}(^3D_2) = 5187.0744 \text{ MeV and } \alpha = 4.0 \text{ Fm}^{-2}. \tag{9}$$

They give the binding energy -1.6964 MeV with accuracy $\varepsilon$ = 10$^{-4}$, what is in very good agreement with experimental value -1.69635 MeV (see Ref. 27), charge radius 2.41 Fm and mass radius 2.33 Fm, AC = 0.033(1) over the range 2 – 12 Fm. In the energy interval from zero up to 1.0 MeV corresponding shift is equal 180°.

It should be noted, this excited state may be associated also with $^5S_2$ wave without FS but with spin channel $S$ = 2. We do not examine this state here, however the corresponding potential was found

$$V_{BS}(^5S_2) = 279.6746 \text{ MeV and } \alpha = 4.0 \text{ Fm}^{-2}. \tag{10}$$

This potential gives the characteristics of the level very close to those are coming from potential (9) binding energy -1.6964 MeV with accuracy $\varepsilon = 10^{-4}$, charge and mass radii 2.42 Fm and 2.45 Fm respectively, but AC = 1.10(1) over the range 2 – 25 Fm. Scattering phase shift is less than 0.1° up to 1.0 MeV. As it is clear these two options give considerably different values for the asymptotic constants. Thus, it is interesting proposal for the experimental measurement of AC (for example, as it was done in Ref. 39) with a view to conclude either this exited level belongs to $^3D_2$ wave or to $^5S_2$.

For the parameters of $^3S_1$ potential without FS corresponding to the third ES of $^{12}$B nucleus ($J^\pi = 1^-$) in $n^{11}$B channel lying at 2.62 MeV the following values have been found

$$V_{BS}(^3S_1) = 266.3015 \text{ MeV and } \alpha = 4.0 \text{ Fm}^{-2}. \qquad (11)$$

They give the binding energy -0.7492 MeV with accuracy $\varepsilon = 10^{-4}$, what coincides with experimental value (see Ref. 27), charge radius 2.43 Fm, and AC equals 1.07(1) over the range 2 – 30 Fm. Fig. 1 shows the corresponding phase shift (dash curve).

For the parameters of $^3P_0$ potential with FS corresponding to the fourth ES of $^{12}$B nucleus ($J^\pi = 0^+$) in $n^{11}$B channel the following values have been found

$$V_{B.S.}(^3P_0) = 3156.9385 \text{ MeV and } \alpha = 4.0 \text{ Fm}^{-2}. \qquad (12)$$

It gives the binding energy -0.6470 MeV with accuracy $\varepsilon = 10^{-4}$, what coincides with experimental value (see Ref. 27), charge radius 2.41 Fm and, AC = 0.25(1) over the range 2 – 24 Fm. In the energy interval from zero up to 1.0 MeV corresponding phase shift is decreasing up to 175°.

## 5. Total Cross Sections

First, the most strong dipole electric $E1$ transition $^3S_1 \xrightarrow{E1} {}^3P_1$ to the GS, i. e. the $(n,\gamma_0)$ process, was treated with zero potential for scattering $^3S_1$ wave and set (6) for the final nucleus. The corresponding cross section is given in Fig. 3 by solid curve. It is seen that within this potential experimental data at 25 meV (closed triangles) and in the energy range 23 – 61 keV (open circles) have been reproduced reasonably well.

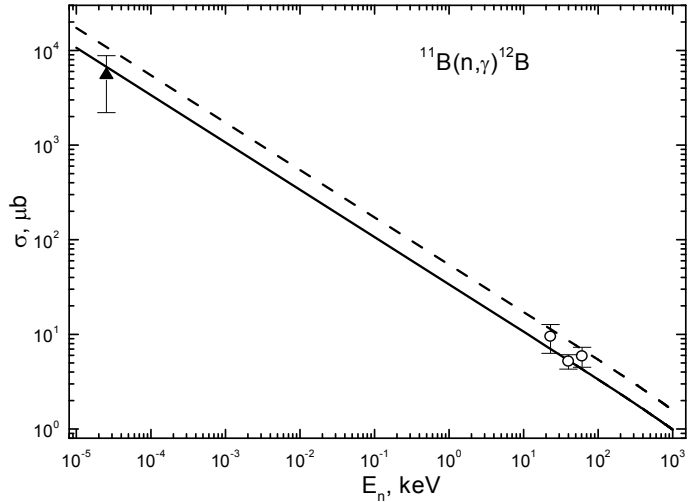

Fig. 3. Total cross sections of $^{11}$B$(n,\gamma_0)^{12}$B reaction.

Experimental data: ▲ – Ref. 41, o – Ref. 42. Curves correspond to the calculation of the total cross sections for the transitions to the GS with potentials given in Sect. 4: potential (6) – solid curve; potential (7) – dash.

For comparison in Fig. 3 by dash curve similar results for the $(n,\gamma_0)$ process are given being obtained with GS potential (7), which gives AC equals 0.55 what coincides with the results of Ref. 39. In this case calculated cross sections more or less describe within the error bars experimental points at 23 – 61 keV,[42] but fail in description of the crucial point at 25 meV[41]. So, we concluded that potential (6) is more appropriate one for the treating of $(n,\gamma_0)$ cross section. Difference in AC of Ref. 39 and those calculated with potential (6) is not very considerable. Note, Ref. 39 was published 35 years ago, so it seems reasonable to refine a value of AC in the future.

Let us turn now to analysis of the total cross sections of $^{11}B(n,\gamma)^{12}B$ reaction in the energy range 1 – 1000 keV and pay special attention to the signature of the resonances at 21 and 430 keV. In Fig.4 we compare the available experimental data and present calculations with account of $E1$ transition amplitudes (a)-(g) listed in Sect. 2.

On the background of $^3S_1 \xrightarrow{E1} {}^3P_1$ transition to the GS (long dash) fist resonance appears at 20.8 keV due to $^3D_3 \xrightarrow{E1} {}^3P_2$ transition from the resonating $^3D_3$ wave to the first ES $^3P_2$ (dot dash). For completeness let us remark that similar transitions from none-resonating $^3D_2$ and $^3D_1$ waves led to the cross sections less than $10^{-3}$ $\mu b$ at 1.0 MeV according our calculations, so they give minor input on the whole.

Transition $^3P_2 \xrightarrow{E1} {}^3S_1$ to the third $^3S_1$ ES from the resonating $^3P_2$ scattering wave is leading to the cross section at 430 keV higher than 1mb (dot curve in Fig. 4). Input of none-resonating partial cross sections $^3P_1 \xrightarrow{E1} {}^3S_1$ and $^3P_0 \xrightarrow{E1} {}^3S_1$ is not essential (dash double-dot curve in Fig. 4).

Partial cross section corresponding to $^3S_1 \xrightarrow{E1} {}^3P_2$ transition to the first ES (densely dots) is parallel to the curve conforming to $^3S_1 \xrightarrow{E1} {}^3P_1$ transition to the GS, and. Transition to the fourth ES $^3S_1 \xrightarrow{E1} {}^3P_0$ have been also calculated, but as it seen in Fig 4 (short dash) it is practically insignificant.

Finally, the total sum of partial cross sections is shown by solid curve in Fig. 4. Experimental data for the transitions to the GS are taken from Ref. 42 (open circles), and data for the sum of partial cross sections are taken from Refs. 42-44.

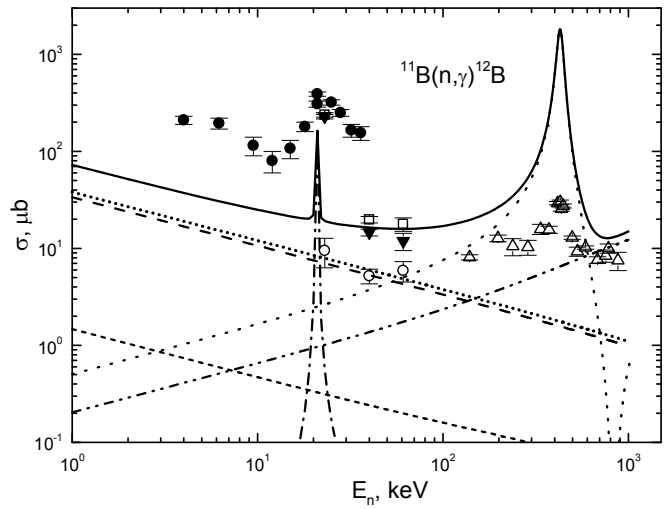

Fig. 4. Total cross sections of $^{11}B(n,\gamma)^{12}B$ reaction in the energy range 1 – 1000 keV.
Including (a) – (g) amplitudes. Experimental data: ▼ – transition to the second ES lying at -1.6964 MeV,
Ref. 42; ● – ref. 43; □ – sum of partial cross sections from Ref. 42; Δ – Ref. 44; o – capture to the GS, Ref.
42. Commentaries for the calculated curves are given in the text.

Let us present another to somewhat extent alternative treatment of $^{11}B(n,\gamma)^{12}B$ reaction in the energy range 1 – 1000 keV illustrated by Fig. 5. There transitions (a), (b), and (f) are preserved with same notations as in Fig. 4. Additional transitions (h) - (*l*) have been examined. So, instead of the process (c1) $^3P_2 \xrightarrow{E1} {}^3S_1$ transition (h) $^3P_2 \xrightarrow{E1} {}^3D_2$ was taken into account (dot curve in Fig. 5). Solid line is the result of the summation of all mentioned partial cross sections.

In Fig. 5 transition $^3P_2 \xrightarrow{M1} {}^3P_1$ which contributes to the energy resonance region at 430 keV is shown by densely dashes, and gives 77.5 $\mu b$ in maximum. For comparison our estimations for $M1$ capture from $^3P_1$ and $^3P_0$ waves with using of potentials (3) give $10^{-2}$ $\mu b$ what is negligibly small.

One more resonating transition $^3P_2 \xrightarrow{M1} {}^3P_2$ is observed in the region of 430 keV with value for the cross section 98 $\mu b$, what is in sum with the last transition amount less than 10% comparing the treated above $E1$ process $^3P_2 \xrightarrow{E1} {}^3D_2$.

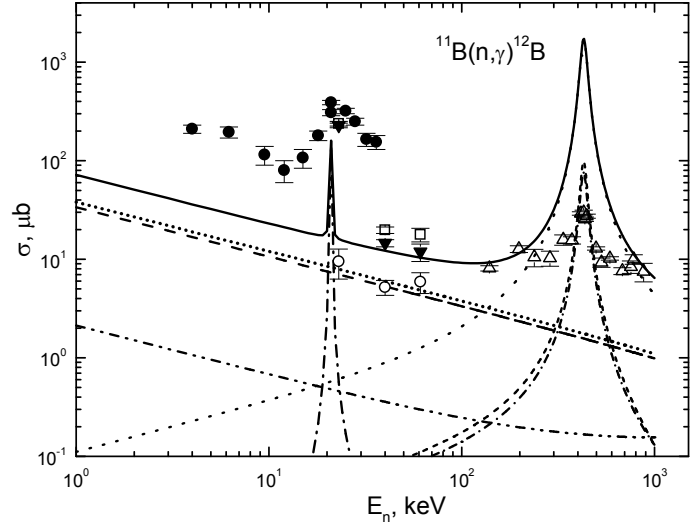

Fig. 5. Total cross sections of $^{11}B(n,\gamma)^{12}B$ reaction in the energy range 1 – 1000 keV.
Including (h) – (l) amplitudes. Experimental data: ▼ – transition to the second ES lying at -1.6964 MeV, Ref. 42; ● – ref. 43; □ – sum of partial cross sections from Ref. 42; Δ – Ref. 44; o – capture to the GS, Ref. 42. Commentaries for the calculated curves are given in the text.

We included also in consideration $^3S_1 \xrightarrow{M1} {^3S_1}$ process (dash double-dot curve in Fig. 5) and found it to be near order less comparing the cross section of $E1$ transition from the same scattering wave to GS.

Resonating capture $^3D_3 \xrightarrow{M1} {^3P_2}$ cross section is near 5 $\mu b$ and does not change the picture at the resonance energy 20.8 keV, as $E1$ process (b) gives for the corresponding cross section a value equals to 144 $\mu b$. In addition all $M1$ transitions from the none- resonating $^3D$ scattering waves to the second ES $^3D_2$ have been evaluated and turned to be less than $10^{-5}$ $\mu b$.

Summarizing the results of Section 5, we may conclude that according Fig. 3 that cross sections at 25 meV and 23 – 61 keV corresponding to the transitions onto GS have been reproduced well and fit experimental data (Refs. 41 and 42). Same statement valid for the cross sections at 21 and 430 keV resonances, but as Fig. 4 shows in the none-resonant energy region experimental data [43] are order of magnitude greater than the calculated ones. There it should be noted, that as AC (this is input information for the construction of interaction potentials) are known for the GS and first ES only, then results obtained for the second, third, and fourth excited states might be regarded as preliminary ones. Moreover, all experimental investigations on this channel (Refs. 42 - 44) were performed in 60-ies of the last century and, apparently, to be specified.

Finally, let us indicate one important predictive opportunity of developed here approach (see also Refs. 6-10). As the calculated cross-section is practically the straight line in the energy range from 10 meV to 10 keV (see solid curve in figure 3), then it may be approximated by the simple function of the form

$$\sigma_{ap}(\mu b) = \frac{33.8364}{\sqrt{E_n(keV)}}$$

The given constant value 33.8364 $\mu b \cdot keV^{1/2}$ has been defined over one point in the cross-sections at the minimal energy equal 10 meV. Relative modulus deviation

$$M(E) = \left| [\sigma_{ap}(E) - \sigma_{theor}(E)] / \sigma_{theor}(E) \right|$$

of the calculated theoretical cross-section ($\sigma_{theor}$) and approximation of this cross-section by the given above function ($\sigma_{ap}$) at the energies up to 10 keV is about 0.3%. If it is assumed, that this form of the energy dependence of the total cross-section will be also conserved at lower energies, then one may estimate the cross-section value which, for example, at the energy 1 $\mu keV$ ($10^{-6}$ eV = $10^{-9}$ keV) is equal to 1.1 barn.

## 6. Conclusion

Within the MPCM based on the deep attractive potentials with forbidden states and coordinated with the spectrum of resonance levels we succeeded to convey properly the behavior of the experimental cross sections of $n^{11}$B radiative capture onto the GS of $^{12}$B in the energy range $10^{-5} – 10^{2}$ keV. Furthermore, within the $E$1 and $M$1 transitions total cross sections have been described at resonance energies as a whole. Potentials for the ground state and first excited are in conformity with basic characteristics of $^{12}$B in $n^{11}$B channel, namely binding energy, charge radius and asymptotic constants.

## ACKNOWLEDGEMENTS

Authors express their heartfelt gratitude to Professor R. Yarmukhamedov for the providing very useful information on the AC in $n^{11}$B channel. This work was supported by the Grant Program No. 0151/GF2 of the Ministry of Education and Science of the Republic of Kazakhstan: The Study of Thermonuclear Processes in the Primordial Nucleosynthesis of the Universe.